# Selective etching of PDMS: etching as a negative tone resist

*S.Z. Szilasi[1,*], L. Juhasz[2]*

[1] *Institute for Nuclear Research, Hungarian Academy of Sciences,*

*H-4001 Debrecen, P.O. Box 51, Hungary*

[2] *University of Debrecen, Dept. of Solid State Physics,*

*H-4010 Debrecen, P.O. Box 2, Hungary*

\* Corresponding author. S.Z. Szilasi

Address: Institute of Nuclear Research of the Hungarian Academy of Sciences

H-4026 Debrecen Bem tér 18/c, Mail: H-4001 Debrecen, POB. 51 Hungary

E-mail address: szilasi.szabi@gmail.com

Tel: +36 52 509 200; Fax: +36 52 416 181

**Abstract**

In this work authors present for the first time how to apply the additive-free, cured PDMS as a negative tone resist material, demonstrate the creation of PDMS microstructures and test the solvent resistivity of the created microstructures.

The PDMS layers were 45 µm and 100 µm thick, the irradiations were done with a focused proton microbeam with various fluences. After irradiation, the samples were etched with sulfuric acid that removed the unirradiated PDMS completely but left those structures intact that received high enough fluences. The etching rate of the unirradiated PDMS was also determined. Those structures that received at least $7.5 \times 10^{15}$ ion × cm$^{-2}$ fluence did not show any signs of degradation even after 19 hours of etching.

As a demonstration, 45 μm and 100 μm tall, high aspect ratio, good quality, undistorted microstructures were created with smooth and vertical sidewalls.

The created microstructures were immersed into numerous solvents and some acids to test their compatibility. It was found that the unirradiated PDMS cannot, while the irradiated PDMS microstructures can resist to chloroform, n-hexane, toluene and sulfuric acid. Hydrogen fluoride etches both the unirradiated and the irradiated PDMS.

1. Introduction

The rapid development in the field of micro/nanofluidics, micro/nanooptics or micro- and nanoelectromechanical systems (MEMS/NEMS) demands the continuous development of lithographic techniques. This includes not only the improvement of the various exposure or irradiation techniques but also the research and development of new resist materials. By the introduction of new resist materials, the quality and/or the dimension of the microstructures may improve and the lithographic processes may become simpler or more reliable.

Poly(dimethylsiloxane) (PDMS) is not unknown in microtechnology. It is widely used mostly as a mold, a casting or replicating material in soft lithography [1] but recent researches showed that it is possible to pattern the polymer with some direct writing techniques also. In 2002, Constantoudis et.al. created structures in liquid, uncured PDMS prepolymer with electron beam lithography and then used the structures as a hard mask. In 2009, Szilasi et.al. irradiated cured PDMS with a high energy focused proton beam and observed significant compaction at the irradiated areas [2]. The compaction effect was applied for the creation of parallel lines with curved surfaces [3] and microlenses [4] in one step, without the need of any further development. In 2011, Tsuchiya et.al. reported that the uncured, liquid phase PDMS polymer crosslinks, thus acts as a negative tone resist if it is exposed to proton irradiation and made microstructures in it [5]. Bowen et.al. created structures by electron beam lithography and studied the change of Young's modulus as a function of the delivered dose in 2012 [6]. In 2016, Gorissen et.al. patterned PDMS through SU8 mask by reactive ion etching (RIE) [7]. Others made the PDMS pre-polymer photosensitive by various additives [8,9].

The application of PDMS as a resist material in direct writing lithography is based on the chemical modification of the polymer due to irradiation. The absorbed radiation creates excited states, ions and free radicals [10] in the polymer that initiate a variety of chemical reactions. The result may be cross-linking, chain scissioning, or the two simultaneously. In cured PDMS, chain scissioning prevails that results in the degradation of the polymer structure. Due to irradiation, the main Si-O-Si chain brakes, functional groups split and the volatile products (e.g. $H_2$, $CH_4$ and $C_2H_6$ gases) leave the irradiated volume [10]. These processes lead to the transformation of the polymer to $SiO_x$ [11] an inorganic, silica-like product. The material properties of the irradiated and degraded PDMS are significantly different from the unirradiated polymer. During irradiation, the initially elastic material becomes hard, rigid and glass-like. Its Young's modulus depends on the irradiation dose and can be varied over approximately seven orders of magnitude [6]. The refractive index can be also tuned by changing the irradiation parameters [12].

Thanks to a range of advantageous properties, it is not surprising that poly(dimethylsiloxane) is probably the most widely used silicon-based, cross-linkable polymer. Besides it is cost effective and easy to use, the cross-linked PDMS is elastic, optically clear, hydrophobic, chemically resistive, stable and inert. These properties make it a good choice in various applications such as microfluidic chips [13, 14], microreactors [15], hydrophobic valves [16], microlenses [17], contact lenses [18], microstamps [19] or even medical implants [20].

Although the presence of PDMS in numerous research fields and applications is significant, up to now it has not been known how to use the cured polymer as a negative tone lithographic resist material. The creation of micro- or nanostructures in cured PDMS has a lot of advantages compared to the lithography in the liquid pre-polymer. The layer thickness of the cured layer can be arbitrary while the thickness of the liquid phase is limited by the flow parameters of the polymer (viscosity, temperature, orientation of the sample). The cured samples need less attention during sample handling, irradiation and storage also because the cured layer protects the created structures from outside impacts before development. Since PDMS is an insulator material, it charges up at the area of irradiation during exposure to charged beams (electron or ion beams). Due to charging, the liquid PDMS layer flows apart making the creation of structures in infinitely thick layers impossible. This problem, of course, does not arise in case of the cured polymer. The above make the creation of arbitrarily tall structures possible, since the height of the structure is only limited by the penetration depth of the used radiation. If a thin conductive layer is necessary during irradiation due to excessive charging, a thin metal layer may simply be evaporated on the top surface of the cured polymer sample. This does not hinder the adhesion between the substrate and the polymer layer and can be either removed or kept in the development process.

In this paper authors present for the first time how to apply the additive-free, cured poly(dimethylsiloxane) as a negative resist material in proton beam writing (PBW), a direct writing lithography technique.

**2. Experimental**

The samples were created by using Sylgard 184 kit from Dow-Corning, the mixing ratio of the base polymer and the curing agent was 10:1. Glass substrates were cut and cleaned in piranha solution ($H_2SO_4:H_2O_2$ - 3:1) for 5 minutes. The PDMS polymer was spin-coated on the glass substrates in 45 μm and 100 μm thicknesses and then baked at 125 °C for 30 minutes.

The samples were irradiated at the nuclear microprobe facility of HAS-ATOMKI, Debrecen, Hungary [21]. The 45 μm and 100 μm thick samples were patterned by 2 MeV and 2.5 MeV protons, respectively. The size of the beam spot was ~2.5 μm × 2.5 μm, the beam current was 1.3 nA. The penetration depth of the different energy protons was calculated by the SRIM [22] code. These calculations showed that the range of the 2 MeV protons is ~85 μm, while that of the 2.5 MeV protons is ~120 μm in PDMS. Since the polymer layers were much thinner than the penetration depths of protons in the corresponding samples, the particles easily penetrate through the resist layer without suffering considerable lateral scattering creating structures with vertical sidewalls.

To test the etching method, two kinds of patterns were created in the samples. The so called fluence test samples consisted of fifteen parallel lines. Each line was numbered and received different fluences in increasing order. In case of one kind of fluence test samples, the fluences ranged from $1.25 \times 10^{15}$ ion × $cm^{-2}$ (2 000 nC×$mm^{-2}$) to $1.88 \times 10^{16}$ ion × $cm^{-2}$ (30 000 nC×$mm^{-2}$) with approximately $1.25 \times 10^{15}$ ion × $cm^{-2}$ (2 000 nC×$mm^{-2}$) increments, while other fluence test samples had better fluence resolution and received fluences between $1.25 \times 10^{15}$ ion × $cm^{-2}$ (2000 nC×$mm^{-2}$) and $5.63 \times 10^{15}$ ion × $cm^{-2}$ (9000 nC×$mm^{-2}$) in $3.13 \times 10^{15}$ ion × $cm^{-2}$ (500 nC×$mm^{-2}$) increments.

The demonstration test samples consisted of various shape microstructures, such as squares, circles, lines with various widths, dot and column matrices. These samples received $1.25 \times 10^{16}$ ion × $cm^{-2}$ (20000 nC×$mm^{-2}$) fluence.

To develop the samples, concentrated sulfuric acid was used. It was found that dilute sulfuric acid does not remove the unirradiated PDMS, so 98% concentration was used. For the best result, the samples were etched for 15 minutes at 35 °C then placed in distilled water for 2 minutes. In case of high aspect ratio structures, intensive stirring of the etchant or the water is not advised because the structures may brake off.

The etched structures were investigated by a Zeiss Axio Imager Optical Microscope and a Hitachi-S4300-CFE scanning electron microscope (SEM).

Since the above mentioned process makes possible the creation of microfluidic elements, it is important to test which solvents the developed microstructures are compatible with. In the framework of the solvent compatibility test two samples were placed in every solvents, a 45 μm thick unirradiated PDMS sample and some developed microstructures. The experiment happened at room temperature, the time duration was 30 minutes. After removing the samples from the solvents they were dried and examined with an optical microscope. Besides organic solvents some acids were also tested. A solvent or an acid was considered compatible with the microstructures or the polymer layer if after 30 minutes no visible changes (whitening, swelling, any degradation, delamination, etc.) could be observed on them.

3. **Results and discussion**

In the first test, a 45 μm thick fluence test sample with parallel lines was etched for 5 minutes. It was found that the sulfuric acid removed only the unirradiated PDMS and did not etch the structures that received high fluences. The lowest fluence line ($1.25 \times 10^{15}$ ion × $cm^{-2}$ or 2 000 nC×$mm^{-2}$) in the test structure disappeared completely but the others remained. The first good quality line that was not damaged by the etchant in 5 minutes was the one that received $5\times10^{15}$ ion × cm$^{-2}$ fluence (8 000 nC×mm$^{-2}$). Below this fluence the quality of the structures decreased with decreasing fluences but above it seemed to be uniform.

The other test sample with smaller fluence steps ($3.13\times10^{15}$ ion × cm$^{-2}$ or 500 nC×mm$^{-2}$) was used to find the fluence threshold of the development with better accuracy. After 5 minutes of etching, it was observed that the quality and integrity of the lines increased steadily with increasing fluences until it reached $4.38\times10^{15}$ ion × cm$^{-2}$ (7000 nC×mm$^{-2}$) above which they were uniform (Figure 1).

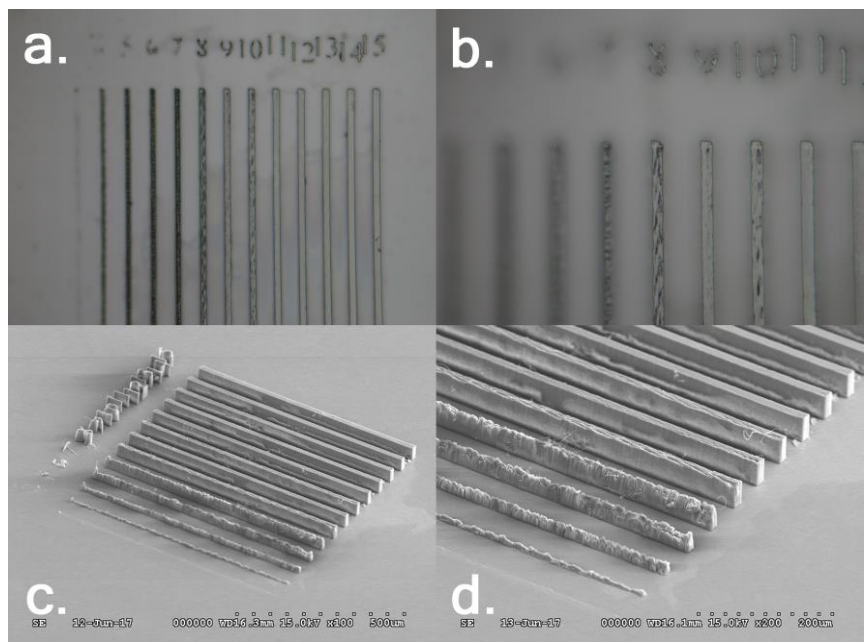

Figure 1. A fluence test sample with $3.13\times10^{15}$ ion × cm$^{-2}$ (500 nC×mm$^{-2}$) resolution after 5 minutes etching. The first line with good quality and integrity is #11 which received $4.38\times10^{15}$ ion × cm$^{-2}$ (7000 nC×mm$^{-2}$) fluence. (a) and (b) are 10x and 20x optical microscope images, (c) and (d) are SEM images taken under 55 degree tilt angle.

To determine how the various fluence microstructures degrade over time in the etchant, the test sample with wider fluence range was placed back into fresh, 98% sulfuric acid. After 20 minutes of etching, the $5\times10^{15}$ ion × cm$^{-2}$ fluence (8 000 nC/mm$^2$) fluence line slightly started to degrade, which becomes obvious at about 50 minutes. After about 3 hours of etching, the $6.25\times10^{15}$ ion × cm$^{-2}$ fluence (10 000 nC/mm$^2$) structure became thinner by 1 μm and showed some signs of degradation. After 19 hours spent in sulfuric acid, some small

7.

cracks could be seen on the edge of this line, and it became thinner by ~1 μm again. Despite the long etching time, all the other structures that received larger fluences remained intact and were in a good condition. This shows that the etching has very high selectivity above this fluence threshold. This concludes that if the creation of microstructures needed with considerable resistance to strong acids, a fluence above $7.5\times10^{15}$ ion × cm$^{-2}$ (12 000 nC/mm$^2$) is needed to be delivered to the structures.

The demonstration test samples could be developed successfully and in a good quality. The etchant cleaned the microstructures well, the cured but unirradiated PDMS was removed completely. The walls were vertical and smooth, the shape of the microstructures were not deformed.

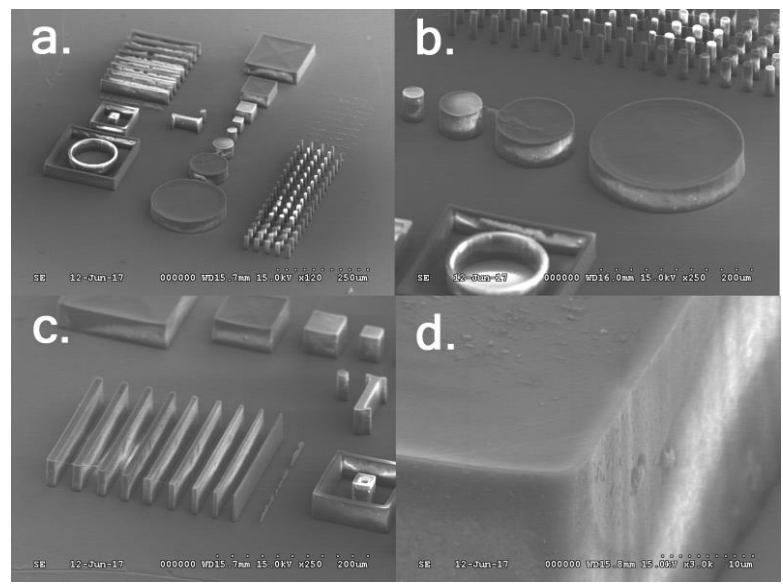

**2. Figure The first 45 μm tall microstructures created by particle irradiation in cured PDMS. Figure (a) - overview image, (b) - various diameter discs, (c) - various width lines, the 2 μm wide one broke off (d) - the edges of a square shape structure – the vertical edge is rounded due to the dose distribution**

Due to the irradiation, the elastic PDMS becomes hard and glass like. Some of the high aspect ratio microstructures broke off due to their rigidity (Figure 2/c), but their adhesion to the glass substrates were very good. The larger structures could not be easily removed from the glass substrate mechanically even by touching them with hand or scratching them with a needle. The explanation for this is probably that the composition and the structure of the highly degraded PDMS is very similar to the glass', so the radicals that formed due to irradiation at the interface attached the two medium together strongly.

When energetic particles penetrate inside a material they suffer scattering. The scattering is more pronounced towards the end of their path where the energy of the ion has already decreased significantly. These irradiations were designed the way that the protons penetrate through the polymer layer and stop inside the glass substrate. This way it can be achieved that the PDMS is modified all the way to the substrate and chemically bonds to it. Since the 100 μm layer thickness is relatively large compared to the 120 μm penetration depth of 2.5 MeV protons in PDMS, the scattering causes visible widening at the bottom of the microstructures (Figure 3.b). The exact height of these microstructures was measured by SEM and it turned out to be 103 μm (Figure 3). The diameter of the narrowest columns was 7 μm at the tip and 15 μm at the bottom. Smaller diameter columns were also irradiated but they broke off from the substrate during etching. This can probably be avoided by further improvement of the development method.

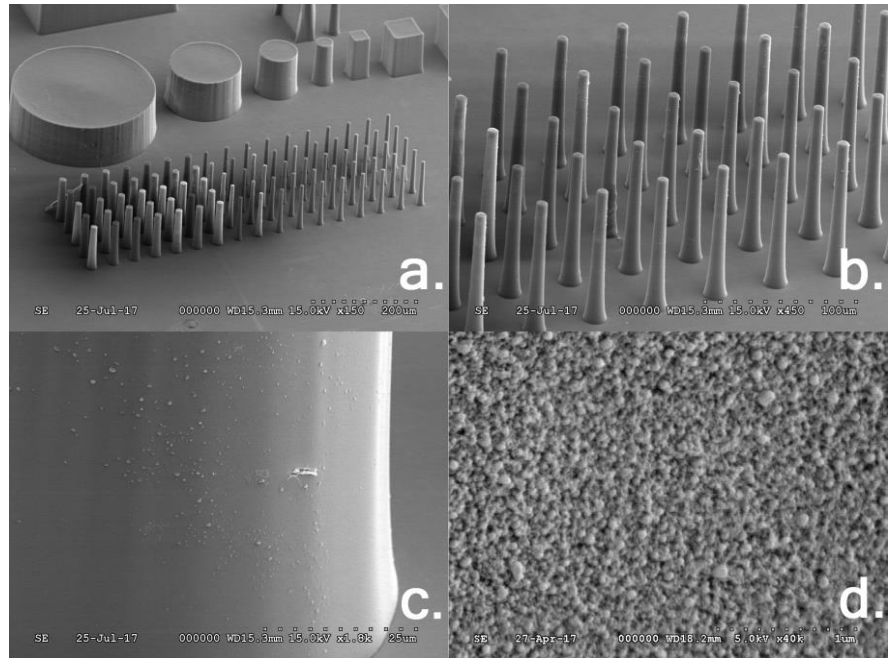

**Figure 3. 103 um tall structures: (a) overview image, (b) the narrowest columns, their diameter was 7 µm at the tip and 15 µm at the bottom, (c) and (d) closeups of the sidewall of a column**

At the development of tall and narrow microstructures, the evaporation of the developing or rinsing liquids may cause problems. If the liquid wets the surface of the microstructures and/or the substrate, and the structures are in contact with the surface of the evaporating liquid then the surface tension and the capillary forces may deform the microstructures or make them collapse. To minimize this effect, the sample has to be kept wet during the development process and the final rinsing liquid has to be non-wetting to the degraded PDMS and the substrate. Since the microstructures become rigid due to irradiation and do not deform easily, they are less sensitive to deformation or collapsing that might occur during drying of the liquids used in the development process.

The average etching rate of the cured, unirradiated PDMS that is immersed into 35 ºC sulfuric acid has also been determined during the experiments and it turned out to be about 0.35 µm/sec. This means that a 45 µm high structure develops in about 2 minutes. However, few microns thin layers etch much faster (~1.5 µm/sec) because the reaction products of PDMS with the sulfuric acid cannot accumulate close to the sample surface and hamper the fresh etchant to reach the microstructures.

The reaction products of PDMS with concentrated sulfuric acid were studied by Lee et al. by IR and mass spectrometry [23]. It turned out that the white precipitate that forms in the reaction consists of low molecular weight oligomers having the structure $(CH_3)_3Si[OSi(CH_3)_2]_xOSi(CH_3)_3$.

The results of the solvent resistivity test (Table 1) showed that the unirradiated PDMS do not, while the irradiated PDMS microstructures do resist to chloroform, n-hexane, toluene and of course sulfuric acid (98%). This is the consequence of changing the material structure of the polymer due to high fluence irradiation. Hydrogen fluoride (38%) etches both the unirradiated and the irradiated PDMS. Previous studies [23] reported that some tested solvents swell PDMS in a significantly longer time and/or at elevated temperatures. At room temperature after the duration of the test, we did not observe the above mentioned effects. Besides the substances listed in Table 1, the effects of 30% potassium hydroxide (KOH) and 30% sodium hydroxide (NaOH) solutions were also tested on irradiated samples. It was found that both solutions etched effectively those areas of the sample that were irradiated with sufficiently high fluences. This means that KOH and NaOH can be used to etch PDMS as a positive resist material. Further results and the details of this study will be published in a separate paper. During the development experiments, it was found that the 30 wt% KOH + 20 wt% IPA + 50 wt% DI water solution at 70 °C temperature etched away both the irradiated and non-irradiated PDMS in 20 minutes. This solution can be used to clean any PDMS residues off of glass wafers.

According to other studies [23], trifluoroacetic acid, dipropylamine and Tetra-n-butylammonium fluoride (TBAF) + tetrahydrofuran (THF) solution also dissolve the cured

PDMS polymer. These substances may also be good candidates to selectively etch the cured and micropatterned PDMS.

| Solvent | Unirradiated PDMS | | Irradiated PDMS microstructures | |
|---|---|---|---|---|
| | Compatible | Not compatible | Compatible | Not compatible |
| Acetic anhydride | Yes | | Yes | |
| Acetone | Yes | | Yes | |
| Acetonitrile | Yes | | Yes | |
| Benzene | Yes | | Yes | |
| Chloroform | | Swells | Yes | |
| Cyclohexane | Yes | | Yes | |
| Dibutyl ether | Yes | | Yes | |
| Diethyl ether | Yes | | Yes | |
| Diethylene glycol monobutyl ether | Yes | | Yes | |
| Ethanolamine | Yes | | Yes | |
| Ethyl alcohol | Yes | | Yes | |
| Hydrochloric acid (37%) | Yes | | Yes | |
| Hydrogen fluoride (38%) | | Etches | | Etches |
| Hydrogen peroxide (30%) | Yes | | Yes | |
| Isopropanol | Yes | | Yes | |
| Methanol | Yes | | Yes | |
| Morpholine | Yes | | Yes | |
| n-Butyl alcohol | Yes | | Yes | |
| n-Hexane | | Swells | Yes | |
| Nitric acid (68%) | Yes | | Yes | |
| Petroleum ether | Yes | | Yes | |
| Sulfuric acid (98%) | | Etches | Yes | |
| tert-Butyl alcohol | Yes | | Yes | |
| Tetrahydrofuran | Yes | | Yes | |
| Toluene | | Swells | Yes | |
| Water | Yes | | Yes | |
| Xylene | Yes | | Yes | |

**1. Table The results of the 30 minute solvent and acid compatibility test**

PDMS is capable of the creation of cured polymer layers with arbitrary thicknesses starting from the nanometre regime. Due to the hardness, chemical resistivity, cost-effectiveness and good adhesion of the micro-/nanostructures, this polymer may be a great choice for hard masks that need to resist the erosion of wet or dry etching in various lithographic processes.

## 4. Conclusions

Authors have found how the additive-free, cured and proton irradiated PDMS can be selectively etched as a negative tone resist material. In this experiment, 45 μm and 100 μm thick cured PDMS layers were irradiated with a focused proton microbeam with various fluences and then etched with 98% sulfuric acid. The etchant removed all the unirradiated PDMS very well while left the irradiated structures intact if the irradiation fluence was high enough. The etching rate of the unirradiated PDMS was about 0.35 μm/sec. In case the irradiation fluence exceeded $7.5 \times 10^{15}$ ion × $cm^{-2}$ (12 000 $nC/mm^2$), no signs of degradation was observable on the structures even after 19 hours of etching in concentrated sulfuric acid. This indicates how high the selectivity of this etching method is. If the creation of microstructures with considerable resistance to strong acids or organic solvents is necessary, at least the above fluence needs to be delivered to the structures.

With this technique, good quality, smooth and vertical sidewall, undistorted, high aspect ratio microstructures were created in 45 μm and 100 μm thick PDMS layers. The microstructures become glass-like, rigid and adhered to the glass substrate very well.

The solvent resistivity of the created microstructures was also tested. It was found that the unirradiated PDMS cannot while the irradiated PDMS microstructures can resist to chloroform, n-hexane, toluene and of course sulfuric acid (98%). Hydrogen fluoride (38%) etches both the unirradiated and the irradiated PDMS.

It was also an important finding that KOH and NaOH solutions could be used to selectively etch PDMS as a positive resist material. These results will be presented in a separate paper.

## 5. Acknowledgements

This work was supported by the National Research, Development and Innovation Fund No. PD 121076, by the Hungarian Scientific Research Fund OTKA No. K 108366 and by the TAMOP 4.2.2.A-11/1/ KONV-2012-0036 project, which is co-financed by the European Union and European Social Fund.

The work was also supported by the GINOP 2.3.2-15-2016-00041 (co-financed by the European Union and the European Regional Development Fund).